\documentclass[a4paper,twocolumn,10pt,unpublished]{quantumarticle}
\pdfoutput=1
\usepackage[utf8]{inputenc}
\usepackage[english]{babel}
\usepackage[T1]{fontenc}
\usepackage{amsmath}
\usepackage{hyperref}
\usepackage{tikz}
\usepackage{lipsum}
\usepackage{float}

\hypersetup{
  citecolor   = blue 
}

\begin{document}

\title{Quantum classical hybrid neural networks for continuous variable prediction}

\author{Prateek Jain}
\affiliation{\textit Fractal Analytics, Gurgaon, India}
\affiliation{\textit Universidad Politécnica de Madrid, Spain}
\email{pratjz@gmail.com}

\author{Alberto Garcia Garcia}
\affiliation{\textit Universidad Politécnica de Madrid, Spain}
\affiliation{\textit Accenture, Madrid, Spain}


\begin{abstract}
Within this decade, quantum computers are predicted to outperform conventional computers in terms of processing power and have a disruptive effect on a variety of business sectors. It is predicted that the financial sector would be one of the first to benefit from quantum computing both in the short and long terms.  In this research work we use Hybrid Quantum Neural networks to present a quantum machine learning approach for Continuous variable prediction.
\end{abstract}

\section{Introduction}
Quantum Machine Learning is an emerging field that shows promising approaches to solve intractable problems. There are quite a few quantum machine learning algorithms that have emerged from their classical counterparts for example the so called Quantum Neural Networks. Quantum neural networks are variational quantum circuits that store quantum information in continuous degrees of freedom, like the amplitudes of an electromagnetic field. Multiple layers of continuously parameterized gates, which are essential to quantum processing, makes up this circuit \cite{cvqnn1}.

The main goal of this research work is to develop a somewhat clearer understanding of the promises and limitations of the current quantum algorithms for machine learning and define some future directions for research. In this work, we will use a parameterized quantum circuit as one of the layer of a Hybrid Quantum Neural Network to present usage for financial applications. The work focuses to demonstrate the use of Hybrid QNNs for Continuous variable predictions for example Asset price prediction in this case. We use the Boston housing data taken from the \href{http://lib.stat.cmu.edu/datasets/}{StatLib library} maintained at Carnegie Mellon University for training the model.

\section{Quantum ML, preliminaries}
In last two decades due to increased computational power and the availability of vast amounts of data, Machine Learning \& Deep Learning \cite{cite2} in particular has seen an immense success with applications ranging from computer vision \cite{cv1} to Natural Speech Synthesis \cite{citenlp1}, playing complex games like Atari \& Go \cite{atari1}. However, over past few years, challenges have surfaced that threaten to slow down this revolution. These challenges are, increasingly overwhelming size of the available data sets \& nearing end of Moore’s law. While novel developments in hardware architectures, such as graphics processing units \href{https://en.wikipedia.org/wiki/Graphics_processing_unit}{GPUs} or tensor processing units \href{https://cloud.google.com/tpu/docs/tpus}{TPUs}, enable orders of magnitude of  improved performance compared to central processing units \href{https://en.wikipedia.org/wiki/Central_processing_unit}{CPUs}, they still can barely cope up much with increasing computational needs.

On the other hand, a new technological paradigm Quantum computing, term first popularised  by famous physicist Richard Feynman \cite{rf1} \cite{cite5} a form of computation that makes use of quantum-mechanical phenomena such as superposition \& entanglement \cite{epr1}, is predicted to overcome these limitations in classical computers. Quantum algorithms, have been investigated since 1980s \cite{shor1}. In recent years one area that has received particular attention is quantum machine learning \cite{cite1} the interplay of quantum mechanics and machine learning.

\subsection{Quantum Neural Networks (QNN)}
Parameterized quantum circuits (PQC) \cite{cite6} are quantum circuits which are primarily made with a combination of fixed gates like CNOT gates \& adjustable gates like Pauli rotations \cite{cite7}. The adjustable parts of the circuit are parameterized, it is these parameters which are optimized to converge to an optimal solution or expected value. Quantum processors are used to evaluate the circuits, while the parameters are optimized using various loss function evaluation techniques like gradient descent on a classical computer. This hybrid approach is much less demanding in terms of number of qubits and the number of layers required in the quantum circuit hence this hybrid approach is much more suitable for NISQ era \cite{cite8}.

Lately, there has been a lot of research on the use of PQC as machine learning models, also commonly termed as Quantum Neural Network (QNN) outlined by Farhi and Neven \cite{cite3} and comprehensive comparison Classical NN vs Quantum NN done by Abbas et al \cite{amira3}. 

QNNs can attribute their origin in discussions for the essential role which quantum processes play in human brain. For example, Roger Penrose has argued that a new physics binding quantum phenomena with general relativity can explain such mental abilities as understanding, awareness and consciousness \cite{penrose1}. 

\begin{figure}[H]
\centering
\includegraphics[scale=0.52]{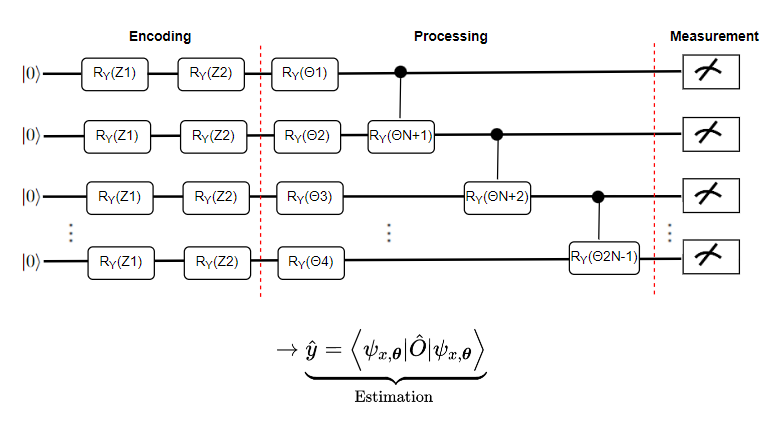}
\caption{General architecture of a QNN.}
\label{fig:figure1}
\end{figure}

In general, typical structure of QNNs can be broadly broken into three stages: feature encoding, processing and measurement with potential post-processing. This general procedure is summarized in Fig.~\ref{fig:figure1}

The architecture consists of three key steps:\\

1. Encode the feature vector $x$ to n-qubits by using a unitary transformation

$$
\left|\psi_x\right\rangle=U_{\phi(x)}|0\rangle
$$

2. Next, a circuit parameterized by parameters $\theta=(\theta 1, \cdots, \theta \mathrm{n})$ is applied which is equivalent to the following initial state
$$
\left|\Psi_{x_a \theta}\right\rangle=U_\theta\left|\psi_x\right\rangle
$$
This circuit $U_\theta$ is commonly called an ansätze \& serves as a general way of transforming the state $\left|\psi_x\right\rangle$ by encoding the data in the Hilbert space. \\

3. Lastly the measurements of the circuit would lead to estimation of the expectation value of the observable $\hat{O}$ this state will be our model output.
$$
\hat{y}=f_{\cos }(x ; \theta)=\left\langle\psi_{x_0 \theta}|\hat{O}| \psi_{x_n \theta}\right\rangle
$$
Now we would need to optimize this output to get suitable results \& thus we will need to adjust the parameters ...$\theta \mathrm{n}$ to minimize some loss function for example
$$
L(\boldsymbol{\theta})=\frac{1}{N} \sum_{i=1}^N L\left(\hat{y}_i, y_i\right)
$$
In the following sections, we will briefly look at few options of feature encoding, circuit optimization \& loss landscape

\subsection{Feature Encoding}
The very first step to initialize a circuit is feature encoding which encodes classical data $x$ into quantum data $|\psi(x\rangle=U(x)|0\rangle \otimes n$ using a parameterized quantum circuit. There are mainly three ways of doing feature encoding i.e. Qubit encoding, Angle encoding \& Amplitude encoding \cite{citelloyd1}.

\subsubsection{Basis Encoding}
Also called as Qubit encoding in this data encoded in qubits as it is encoded in bits thus it is the simplest method, let the classical dataset be $\text { data }=\left\{x_1, x_2, \ldots, x_N\right\}$ where each data point $x_i$ has $f$ feature variables, let each data point $x_i$ be denoted by a unique bitstring $\vec{b}^i=b_0^i b_1^i \ldots b_{l-1}^i$, where $l$ depends on the data type, if it is 32-bit floating point number then $l=32 \mathrm{f}$ for $f$ features, then the data point $x_i$ can be encoded in quantum register consisting of $l$ qubits:
$$
x_i \mapsto\left|b_0^i\right\rangle\left|b_1^i\right\rangle \ldots\left|b_{l-1}^i\right\rangle \equiv\left|\vec{b}^i\right\rangle .
$$

for example we have 3 features as $x_1=1$, $x_2=2$, $x_2=3$ their binary representations would be  $x_1=001$, $x_2=010$, $x_2=011$ thus corresponding basis encoding uses 3 qubits as $|x_1\rangle=|001\rangle$, $|x_2\rangle=|010\rangle$, $|x_2\rangle=|011\rangle$ resulting in encoded state

$$
|D\rangle = \frac1{\sqrt3}\left(|001\rangle + |010\rangle + |011\rangle\right)
$$
For low depth circuits qubit encoding is suitable, But the state resulting from qubit encoding is very simple mathematically, which might limit the expressive power of the model.

\subsubsection{Angle Encoding}
In angle encoding each feature variable is encoded to a qubit hence will need n qubits, for n features. The encoding is done by performing a Pauli-rotation on each qubit with a rotational angle equal to the corresponding feature.


With Rz gate a Hadamard gate is used on each qubit to create a superposition else these rotations would leave $|0\rangle$ invariant. For example, two features $x = (x_1, x_2)$ can be qubit encoded onto two qubits in the following way using Ry rotations:
$$\small 
R_y\left(x_1\right) \otimes R_y\left(x_2\right)(|0\rangle \otimes|0\rangle)=$$
$$\left(\cos \left(\frac{x_1}{2}\right)|0\rangle+\sin \left(\frac{x_1}{2}\right)|1\rangle\right)\otimes\left(\cos \left(\frac{x_2}{2}\right)|0\rangle+\sin \left(\frac{x_2}{2}\right)|1\rangle\right)
$$

Angle encoding produces much more complex feature maps which can be made more complex by repeating the encoding process multiple times. The number of these repeated layers is called the depth of the feature map. However, angle encoding is also much more computationally expensive, since it requires circuit depth of $O(n^2)$ for n features and fully connected qubits.

\subsubsection{Amplitude Encoding}
In Amplitude encoding features encoded as amplitudes of a quantum state. Given a data set $x = (x_1,...., x_N)$, where $N = 2^n$, amplitude encoding involves preparing a state $$\left|\psi_{\boldsymbol{x}}\right\rangle=\sum^N{ }_{i=1} x_i|i\rangle$$
on n qubits. Because of the normalization of the quantum states, it is essential to scale the data to make sure that sum of probabilities i.e. $\sum_{i=1}^N\left|x_i\right|^2=1$ holds.\\\\
for example let say we have a datapoint $x$ with features $(1, 0, 6.8, 1.0)$ then $x_{norm} = \frac1{\sqrt{48.24}}(1, 0, 6.8, 1.0)$, the corresponding amplitude encoding uses two qubits resulting in encoded state as

$$|\psi_{x_{norm}}\rangle = \frac1{\sqrt{48.24}}\left(1|00\rangle + 0|01\rangle + 6.8|10\rangle +1.0|10\rangle \right)$$

$$\;\; = \frac1{\sqrt{48.24}}\left(1|00\rangle + 6.8|10\rangle +1.0|10\rangle \right)$$


The key advantage of amplitude encoding is that the number of amplitudes is pretty much limitless in the number of qubits therefore only $log_2(n)$ qubits are needed to encode n features thus an enormous amount of information can be encoded with each qubit added. However, there is not much eﬃcient way of preparing the state.

\subsection{Optimization of a PQC/QNN}
A key step for hybrid PQCs is the optimization of the parameters $\theta$ for the initial ansatz. These parameters are optimized with respect to an objective function specific to a given problem. There are multiple methods for optimizing PQCs, one example of such method is numerical diﬀerentiation of the loss function:
{\small $$
\frac{\partial}{\partial \theta_i} L(\theta) \approx \frac{L\left(\theta_1, \cdots, \theta_i+\epsilon, \cdots \theta_{n_\theta}\right)-L\left(\theta_1, \cdots, \theta_i, \cdots \theta_{n_\theta}\right)}{\epsilon}
$$}
One can optimize the parameters using techniques similar to gradient descent. 

\subsection{Barren Plateaus in Loss Landscape}
While recent researches have shown multiple promising characteristics of QNNs, like faster training \& better generalizeability but these studies have been largely focused on smaller systems. McClean et al \cite{barren1}. established results relating the magnitude of the gradient to the number of qubits. They found that the variance of the expected value of randomly initialized PQCs vanish exponentially with increasing number of qubits \& circuit depth.

The vanishing of PQCs gradients manifests itself as loss landscapes that are extremely flat in most of parameter space, hence the name Barren Plateaus, similar to the vanishing gradient phenomenon of classical neural networks as the depth increases. This is currently a very active research area few techniques have been proposed like each layerwise learning of gradient for the parameters \cite{layerwise1}

To evaluate an ansatz design different techniques \& circuit descriptors are used such as the circuit expressability which is the efficiency with which a quantum circuit may exploit the Hilbert Space, and entanglement capability which quantifies a circuit’s capability of detecting correlations among features.

\subsection{Expressability}
A circuit is considered to be expressive if it is able to generate pure quantum states that are a good representation of the Hilbert space in consideration. Sim et al (2019)'s \cite{sim1} method compares the ensemble of Haar random states to distribution of states read by sampling a Circuit's(PQC's) parameters to calculate expressability they proposed to approximate the distribution of fidelities as the overlap of states defined as:
$$
F=\left\langle\psi_\theta \psi_\phi \mid \psi_\theta \psi_\phi\right\rangle^2
$$
For the ensemble of Haar random states the probability density function of fidelities is defined as:
$$
P_{\text {Harr }}(F)=(N-1)(1-F)^{N-1},
$$
where $\mathrm{F}$ corresponds to the fidelity and $\mathrm{N}$ is the dimension of the Hilbert space \cite{sommers1}. After collecting sufficient samples of the state fidelities, the Kullback-Leibler (KL) divergence \cite{kl1}, between the estimated fidelity distribution and that of the Haar distributed ensemble can be computed to define expressibility as:
$$
\text { Expr }=D_{K L}\left(\left(\hat{P}_{P Q C}(F ; \theta) \| P_{\text {Haar }}(F)\right),\right.
$$
lower the KL divergence higher is the Expressability of the circuit.

\subsection{Entangling Capability}
A parametrized quantum circuit with lesser number of layers which is able to generate highly entangled states, is said to have better advantage to capture nontrivial correlation in quantum data Schuld and Petruccione \cite{cite1}. Strongly entangled circuits can be created by repeating circuit layers made of various two-qubit parameterized gates like CNOT, CZ. Sim et al (2019) \cite{sim1} proposed using Meyer-Wallach (MW) \cite{meyer1} entanglement measure Q to approximate entangling capability of a PQC. For a given PQC the entangling capability(Ent) can be estimated by sampling the circuit parameters and calculating the average of the MW measure Q of output states defined as:
$$
E n t=\frac{1}{S} \sum_{\theta_i \in S} Q\left(\psi_{\theta_i}\right),
$$
where $S$ denotes the set of sampled circuit parameter vector $\theta$.

\subsection{QML for Finance Applications}
A myriad of problems in finance industry are addressed using Machine learning \& deep Learning techniques but there are still a lot of cases where classical computation techniques turn out to be inadequate for example NP-hard problems like portfolio optimization \cite{qmlfinance2}, currency arbitration \cite{qmlfinance3} etc..There have been a lot of areas identified \cite{qmlfinance1} where-in proposed QML techniques could be used to address such problems. For this study we chose asset pricing use case wherein we use QML techniques to predict property prices.

\section{Frameworks \& libraries}
\subsection{PennyLane}
Is an open-source framework built around quantum differentiable programming by Xanadu implemented in Python to achieve machine learning tasks with quantum computers \cite{pennylane1}. It Supports hybrid quantum and classical models allowing users to connect quantum hardware with PyTorch, TensorFlow, Qiskit, Cirq etc..therefore it is Hardware agnostic i.e. same quantum circuit model can execute on different backends and allows plugins for access to diverse devices, including Strawberry Fields, Amazon Braket, IBM Q, Google Cirq, Rigetti Forest, Microsoft QDK..

\subsection{StrawberryFields}
Strawberry Fields is an open-source framework for photonic quantum computing again by Xanadu \cite{strawberry1}. In particular, Strawberry Fields allows to Construct and simulate continuous-variable quantum photonic circuits. Provided simulators include highly optimized Gaussian, Fock, and Bosonic numeric backends and a TensorFlow backend for backpropagation.

Compared to qubit-based systems, photonic quantum programs use a different gate set, and have different near-term applications. Strawberry Fields is a full-stack solution for constructing, compiling, simulating, optimizing, and executing photonic algorithms.

\subsection{Tensorflow Quantum \& Keras}
is another framework by google it brings the power of Tensorflow to the quantum world\cite{tfq1}. It provides high-level abstractions for the design and training of QNNs including discriminative and generative quantum models under TensorFlow and supports high-performance quantum circuit simulators along with high level wrapers for Keras \cite{keras1}.\\

Combination of all these three libraries were used in this work to create Hybrid Quantum Neural networks for predicting property prices from the Boston Housing dataset.

\section{Methodology and Setup}
In this section, we will go through details of implementation of algorithms and architecture designed. The framework is also capable of implementing hybrid models mixing both DNN and QNN layers.

\subsection{DataSet}
We use Boston Housing Data it was originally hosted on UCI Machine Learning Repository but now can be directly accessed from keras.datasets. Data Samples contain 13 feature variables of houses at different locations around the Boston suburbs in the late 1970s. Target or predictor variable is the median price value of the houses (in k\$). The dataset has total 506 samples. Following is description of the features:
.\\\\
CRIM    - per capita crime rate by town\\
ZN      - proportion of residential land zoned for lots over 25,000 sq.ft\\
INDUS   - proportion of non-retail business acres per town\\
CHAS    - Charles River dummy variable (= 1 if tract bounds river; 0 otherwise)\\
NOX     - nitric oxides concentration (parts per 10 million)\\
RM      - average number of rooms per dwelling\\
AGE     - proportion of owner-occupied units built prior to 1940\\
DIS     - weighted distances to five Boston employment centres\\
RAD     - index of accessibility to radial highways\\
TAX     - full-value property-tax rate per 10,000\\
PTRATIO - pupil-teacher ratio by town\\
B       - 1000(Bk - 0.63)\^{}2 where Bk is the proportion of blacks by town\\
LSTAT   - \% lower status of the population\\
MEDV    - Median value of owner-occupied homes in \$1000's, Target Value (Y label)

\subsubsection{Correlation among Features}
Here we create features correlation heatmap to see How each feature is correlated to the target variable MEDV

\begin{figure}[H]
\centering
\includegraphics[scale=0.4, angle =90]{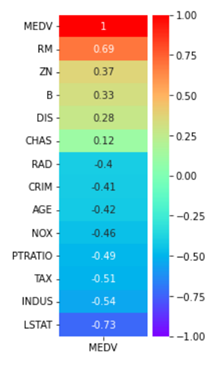}
\caption{Feature correlation Heatmap}
\label{fig:figure6}
\end{figure}

From the correlation heatmap we can infer the importance of features and some insights for example:
\begin{itemize}
\item RM: For a higher RM, one would expect to observe a higher MEDV.This is because more rooms would imply more space hence the cost.
\item LSTAT: For a higher LSTAT, one would expect to observe a lower MEDV. Generally, an area with more so called 'lower class' citizens would have lower demand, hence lower prices.
\item PTRATIO: The prices of houses around public schools are generally lower than those around private schools because there would be a lower teacher-to-student ratio in public schools resulting in less attention dedicated to each student that may impair their performance. Hence one would expect a lower price given a high student-to-teacher ratio due to a lower demand for houses in such areas.
\end{itemize}

We standardize the data by scaling it and removing the mean and variance to address the problem of the data being on different scales.

\subsubsection{Principal Component Analysis (PCA)}

Principal component analysis is done to find the least number of features required to have a good model hence we will reduce number of features and determine the approximate number of qubits required.

\begin{figure}[H]
\centering
\includegraphics[scale=0.34]{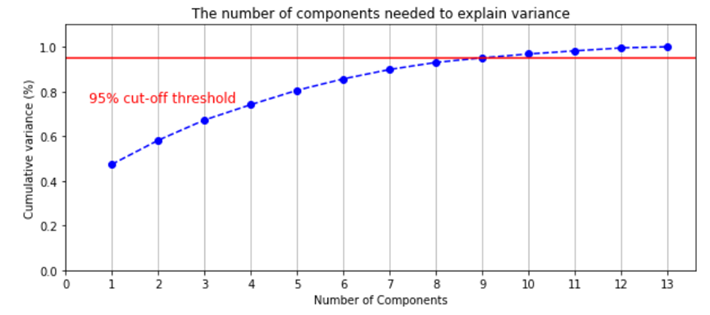}
\caption{Principal Component Analysis (PCA)}
\label{fig:figure7}
\end{figure}

Here we can see that to get at-least 95\% of variance explained we need 9 principal components.

Now since we have our data pre-processed and ready, in the following sections we will employ various different types of Quantum Hybrid \& Classical Neural networks to the dataset for price predictions and compare the performance of all the approaches.

\subsection{Quantum Classical Hybrid Neural Network Architecture}

First step is to create a quantum device, after trying various simulators available in pennylane we selected the \textit{default.qubit.tf} device which is based on tensorflow and has intrinsic support for tensor calculus \& differentiation it is faster than default and many other devices.Then create the pennylane qnode using qnode decorator. 

\begin{figure}[H]
\centering
\includegraphics[scale=0.31]{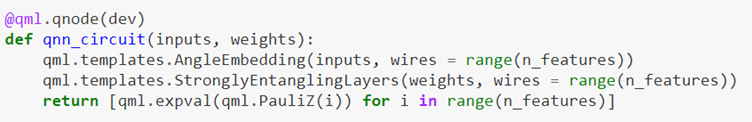}
\caption{Qnode defining the QNN to be executed on device dev.}
\label{fig:figure9}
\end{figure}

Qnode denotes the quantum circuit annotated by the device on which it should execute. This functionality provides immense flexibility in defining quantum circuits allowing multiple devices to be used in parallel for multiple qnodes Fig.~\ref{fig:figure9}, also the device decorator has many other attributes like ‘parallel’ as applicable by the supporting device.In qnode a quantum function is defined which is basically the quantum circuit, we can use various kinds of operations, embedding, circuit templates in this function, we can also define our own gate-based operations \& custom circuit, where in the function takes inputs \& weights as parameters.

In the qnode first step is to map Qubits to features. For which \href{https://docs.pennylane.ai/en/stable/code/api/pennylane.AngleEmbedding.html}{AngleEmbedding} from pennylane is used. It encodes N features into rotation angles of n qubits where $N \neq n$. AngleEmbedding chosen over AmplitutedEmbedding because the dataset is fairly large with 500+ samples and so is the number of features which would be very complex and thus difficult to encode \& run on a simulator or five qubit IBMQ QPU. Wherein AngleEmbedding provides much more complexity \& flexibility than the simpler \href{https://pennylane.ai/qml/glossary/quantum_embedding.html}{BasisEmbedding} and relatively easy to use compared to AmplitutdEmbedding.

For QNN layer StronglyEntanglingLayers template is used, it will create a circuit with layers consisting of single-qubit rotations and entanglers, inspired by the circuit-centric classifier design described in Schuld et al \cite{cvqnn1} It allows to train Quantum Layer using features as angle parameters and is better suited for predictions involving continuous variables.

Keras is used to create the classical neural network architecture, The first layer clayer\_in defines the input layer \& clayer\_out for predicted output Dense Layer with linear activation is used. The Complete neural network model has the quantum circuit layer sandwiched between the input and output classical NN layers i.e. [clayer\_in, qlayer, clayer\_out]

After creating the qml.qnn.KerasLayer, which simply wraps a QNode into a layer that’s compatible with TensorFlow and Keras, TensorFlow is able to classically optimize the network. The gradient for the quantum part of the network is supplied by the QNode and is calculated by different means depending on the device used (in the strawberryfields.fock case it’s calculated by finite differences), while all other gradients are calculated classically by TensorFlow for example using SGD.

Since linear regression is the problem at hand, selected loss type is mean squared error MSE, it squares the difference before summing them all instead of using the absolute value.

Following is a diagramatic presentation of the entire neural network (input is a 9-dimension feature vector)

\begin{figure}[H]
\centering
\includegraphics[scale=0.31]{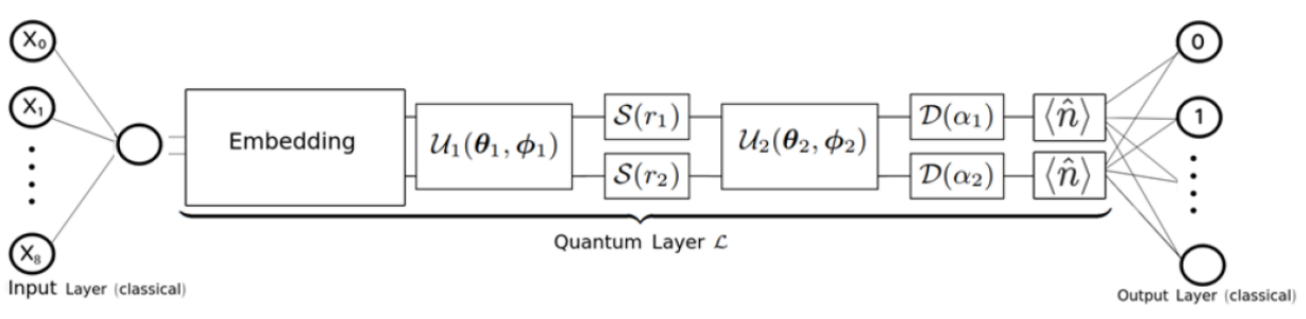}
\caption{Quantum classical Hybrid QNN with one Quantum circuit layer sandwiched between two classical input \& output NN layers }
\label{fig:figure10}
\end{figure}

\subsubsection{Classical Neural Network Architecture}
Classical Neural Network is also created using tensorflow and Keras, the architecture of the NN contains three layers similar to the QNN-Hybrid approach the input layer has number of features as inputs \& one hidden layer followed by an output layer,	Same SGD optimizer is used to reduce the loss. Most of the architecture \& parameters have been kept same as the Hybrid-QNN for fair comparison

\subsubsection{Photonic Quantum Classical Hybrid Neural Network Architecture}
Architecturally this QNN similar to the Hybrid-QNN defined earlier, the key difference is that quantum circuit is built to be executed on a photonic QPU using strawberryfields library.The Qnode (quantum circuit)is created by using DisplacementEmbedding for feature encoding and CVNeuralNetLayers as the photonic quantum neural network layer for continuous variables.

\begin{figure}[H]
\centering
\includegraphics[scale=0.28]{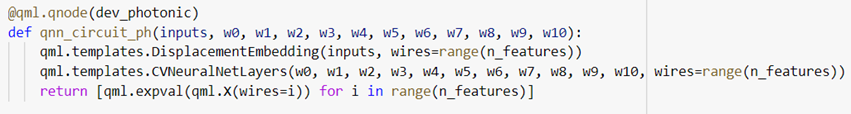}
\caption{Photonic QNN Qnode}
\label{fig:figure11}
\end{figure}

DisplacementEmbedding Encodes N features into the displacement amplitudes r or phases $\Phi$ of M modes(qubits), where $N \leq M$. CVNeuralNetLayers is A sequence of layers of a continuous-variable quantum neural network, as specified in \cite{cvqnn1}. The layer consists of interferometers, displacement and squeezing gates mimicking the linear transformation of a neural network in the x-basis of the quantum system, and uses a Kerr gate to introduce a ‘quantum’ nonlinearity.

Similar to earlier QNN the photonic quantum classical hybrid neural network model has the quantum circuit layer sandwiched between the input and output classical NN layers [clayer\_in, qlayer, clayer\_out]. Here too Stochastic Gradient Descent Optimizer is used.

Essentially three neural network models were trained to predict the house prices, configuration details of these architectures are listed here:

\begin{table}[H]
\small \addtolength{\tabcolsep}{-4pt}

\begin{tabular}{|l|l|l|l|}
\hline
                       & \textbf{Hybrid QNN} & \textbf{Classical NN} & \textbf{Photonic QNN} \\ \hline
\textbf{Features used} & 9                   & 9                     & 3                     \\ \hline
\textbf{Epochs}        & 25                  & 25                    & 25                    \\ \hline
\textbf{Learning Rate} & 0.08                & 0.08                  & 0.08                  \\ \hline
\textbf{Batch Size}    & 5                   & 5                     & 5                     \\ \hline
\textbf{Shots}         & 1024                & NA                    & 1024                  \\ \hline
\textbf{NN layers}     & 3                   & 3                     & 3                     \\ \hline
\end{tabular}
\end{table}

\section{Results}
A machine with core i7 CPU \& 32gb memory was used to conduct the simulations. In the above table we see that the Photonic QNN is using only three feature this is because the photonic library Strawberry Fields Fock backend is very computationally intensive, the memory required for simulation scales like $D^N$, for D dimensions and N wires/qubits. While CV neural nets are conceptually very compelling, they are extremely costly to simulate (because of their infinite-dimensional Hilbert spaces). At 6 features the model error-ed out giving out of memory and at 5 features the model went on training for over 10 hours and then timed out thus for the simulation purpose finally 3 features were used for training, which took around two hours to train \& as expected the model pretty much did not learn anything.
\subsection{Investigating the Loss Landscape}

Figure.~\ref{fig:figure13} illustrates the loss decay comparison with number of epochs, lower the loss higher is the prediction accuracy of the model. We can see that. First Quantum Neural network model performs slightly better than the simple classical neural network with the same settings \& configurations though it took about three times the time to train compared to the classical NN in future with large scale Quantum Computers this could be explored on actual QPUs, the accuracy of the simple NN can be made better with changing the Neural network architecture but for fair comparability the architecture was kpet almost similar. We can see that the QNN is already performing slightly better, this could further be tweaked explored by using different settings like encodings or different ansatz.

Photonic QNN even with very less number of features is also able to perform with near comparable performance demonstrating similar loss decay \& bearing in mind that these are simplest QNNs, In future when higher end Quantum devices are available more complex robust QNN-Hybrid architectures could provide manifold advantages.

\subsection{Comparing Actual vs Predicted}

\subsubsection{Comparing Prices predicted by Hybrid-QNN}

Here we can see in Fig.~\ref{fig:figure14} that the model has generalized fairly well and the predicted prices are having closer variability to the predicted prices, these are results from a very simple QNN trained on a simulator which could be made much better with more complex larger QNN architectures.

\subsubsection{Comparing Prices predicted by Classical NN}
Here we can see in Fig.~\ref{fig:figure15} that the model has not generalized as well as the Hybrid-QNN model, well of course the model architecture could be made much better with current ML techniques \& frameworks to make the predictions most accurate but with comparable settings with the QNN, the QNN generalizes fair amount better.

\subsubsection{Comparing Prices predicted by Photonic QNN}
In this case we notice that the model has not learnt anything at all ref Fig.~\ref{fig:figure16}, this is primarily because of the fact that the number of features used to train the model was very small due to memory and compute power limitations. as stated earlier with the availability of higher end quantum devices this CVQNN architecture could be improved far more. Now with cloud availability of Xanadu's Borealis much better experiments can be tried.

\subsection{Discussion \& Future Work}

From the results and comparisons with fair confidence we can deduce that even with the simplest settings and configurations the quantum parameterized circuits or the so-called Quantum Neural networks generalised fairly well and as the technology and research matures quantum hybrid approaches show promising results over simple classical approaches in many areas especially where there is involvement of classically intractable problems for example Finance \& quantum chemistry wherein plenty of the problems have extremely high dimensionality.

Qubit-based quantum computers have a drawback that they are not completely continuous, since the measurements of qubit-based circuits are mostly discrete. Therefore they are not very suited to continuous variable problems \cite{cite25}. On the other hand the quantum computing architecture which is more suitable to continuous-variable (CV) problems is photonic QPUs where-in the Quantum information is not encoded in qubits, but is encoded in quantum states of continuous spectrum fields like electromagnetic waves. 

Therefore as Future Research, different photonic CV QNN ansatz \& approaches can be explored like some described in \cite{cvqnn1} with actual photonic quantum hardware’s like \href{https://www.xanadu.ai/products/borealis/}{Xanadu's Borealis}.

\section{Acknowledgment}
P. J. is grateful to A. G. G. for his guidance \& suggestions \& to \textit{Universidad Politecnica de Madrid} for supporting this research. The authors acknowledge the use of libraries Pennylane \& Strawberryfield by Xanadu and Tensorflow \& Keras by Google.

\bibliographystyle{quantum}

\onecolumn
\appendix
\section{Charts from the Results}

\begin{figure}[H]
\centering
\includegraphics[scale=0.4]{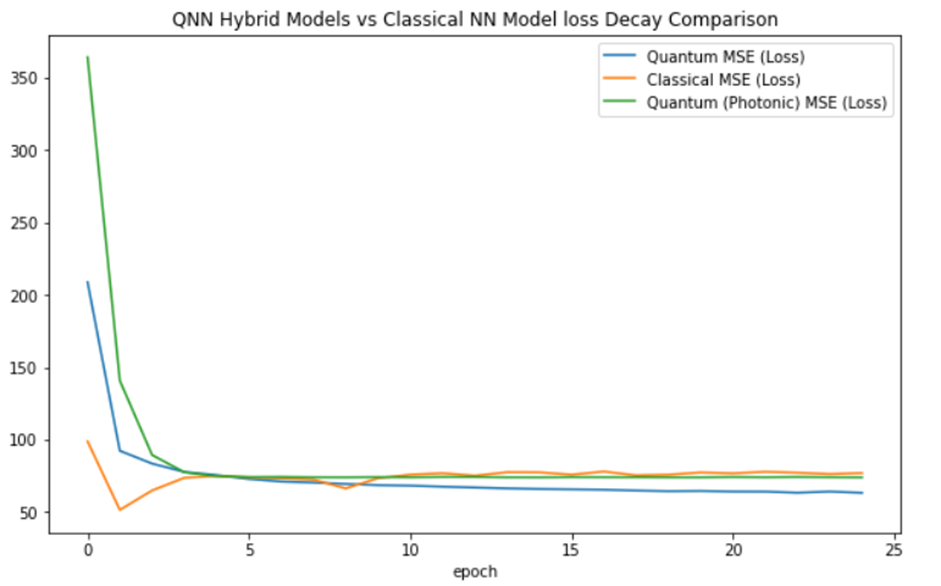}
\caption{Loss Landscape}
\label{fig:figure13}
\end{figure}

\begin{figure}[H]
\centering
\includegraphics[scale=0.4]{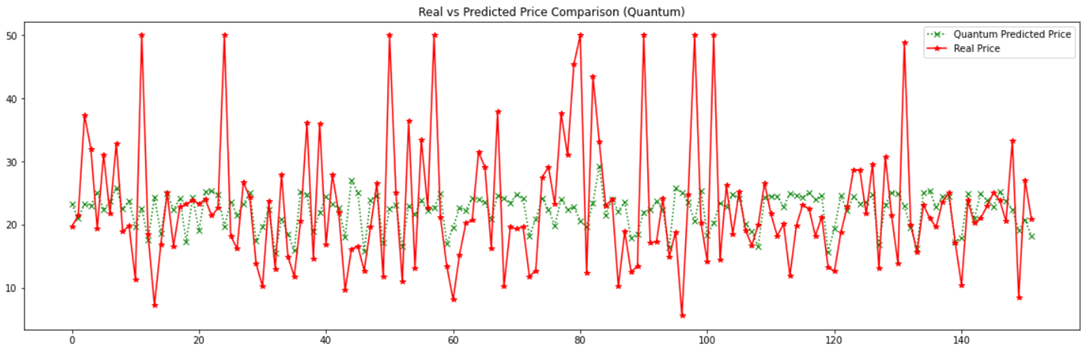}
\caption{Comparing Prices predicted by Hybrid-QNN model}
\label{fig:figure14}
\end{figure}

\begin{figure}[H]
\centering
\includegraphics[scale=0.45]{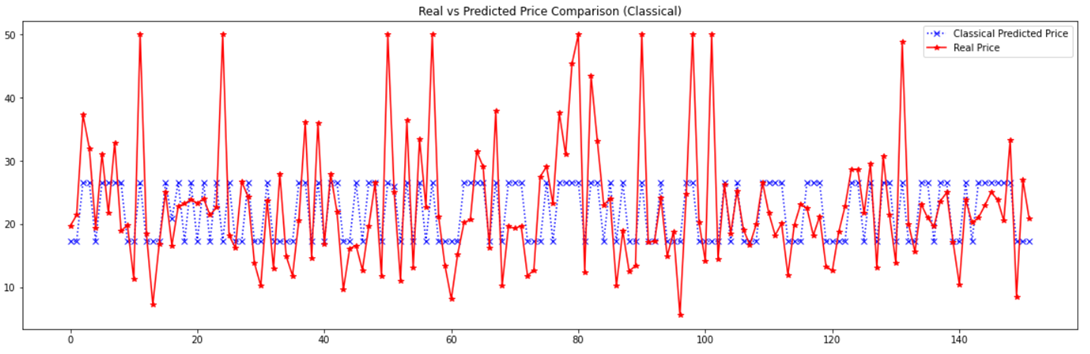}
\caption{Comparing Prices predicted by Classical NN model}
\label{fig:figure15}
\end{figure}

\begin{figure}[H]
\centering
\includegraphics[scale=0.45]{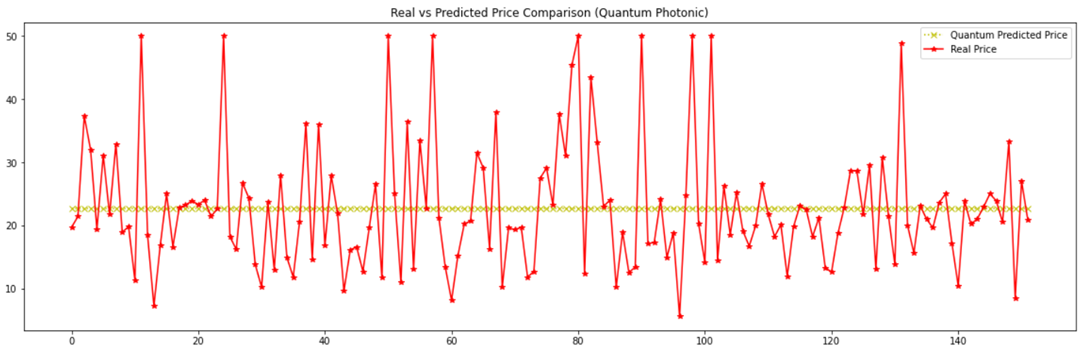}
\caption{Comparing Prices predicted by Photonic QNN model}
\label{fig:figure16}
\end{figure}

\end{document}